# Comment on "Two-dimensional porous graphitic carbon nitride $C_6N_7$ monolayer: First-principles calculations [Appl. Phys. Lett. 2021, 119, 142102]"

Bohayra Mortazavi[a,*], Fazel Shojaei[b] and and Masoud Shahrokhi[c]

[a]*Department of Mathematics and Physics, Leibniz Universität Hannover, 30167 Hannover, Germany.*
[b]*Faculty of Nano and Bioscience and Technology, Persian Gulf University, Bushehr 75169, Iran.*
[c]*Young Researchers Club, Kermanshah Branch, Islamic Azad University, Kermanshah, Iran.*

*Corresponding author: bohayra.mortazavi@gmail.com;

Recently, Bafekry et al. [*Appl. Phys. Lett. 119, 142102* (**2021**)] reported their density functional theory (DFT) results on the elastic constants of a novel $C_6N_7$ monolayer. They predicted a very soft elastic modulus of 36.29 GPa for the $C_6N_7$ monolayer, which is remarkably low for carbon-nitride 2D lattices. Using DFT calculations, we predict a remarkably higher elastic modulus of 267 GPa for this monolayer. The maximum tensile strength is also predicted to be 20.5 GPa, revealing the outstanding mechanical properties of the $C_6N_7$ monolayer.

We first show the crystal structure of the $C_6N_7$ monolayer, which is illustrated in Fig. 1. DFT calculations are performed by employing the *Vienna Ab-initio Simulation Package* [1,2]. The generalized gradient approximation (GGA) is adopted with the Perdew-Burke-Ernzerhof (PBE) exchange–correlation functional. The plane wave and self-consistent loop cutoff energies are set to be 500 and $10^{-4}$ eV, respectively. To obtain the geometry optimized and stress-free structures, atomic positions and lattice dimension are relaxed using the conjugate gradient algorithm until Hellman-Feynman forces drop below 0.01 eV/Å using a 3×3×1 Monkhorst-Pack [3] K-point grid. The periodic boundary conditions are considered in all directions with a 14 Å vacuum distance along the monolayer's thickness. The hexagonal lattice constant of the $C_6N_7$ monolayer is found to be 11.720 Å, in a close agreement with that reported in the work by Bafekry *et al.*[4].

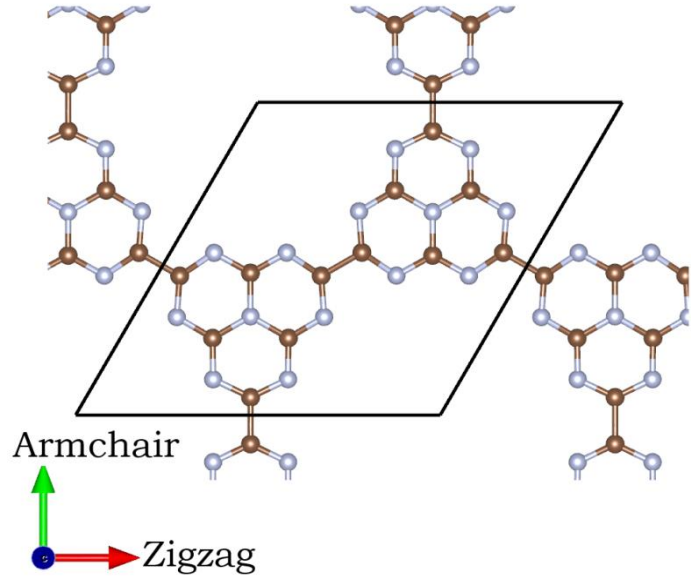

**Fig. 1**, Top view of the $C_6N_7$ monolayer with a fully-planar atomic lattice. In this figure, light blue and brown colors indicate N and C atoms, respectively.

Bafekry et al.[4] predicted the Young's modulus of the $C_6N_7$ monolayer as 36.29 GPa. Using the DFT method and with assuming a thickness of 3.35 Å for the $C_6N_7$ monolayer on the basis of graphene's thickness, the $C_{11}$ and $C_{12}$ of the $C_6N_7$ monolayer are predicted to be 286 and 73 GPa, respectively, equivalent with a Young's modulus of 267 GPa, more than seven folds higher than that predicted by Bafekry et al.[4]. Moreover, they predicted the $C_{13}=C_{23}=2.49$ GPa and $C_{33}=9.05$ GPa for the $C_6N_7$ monolayer. We note that a fully-flat monolayer that is in contact with vacuum on the both sides and is free to move toward the out-of-plane direction upon the geometry optimization, cannot exhibit non-zero $C_{13}$, $C_{23}$ and $C_{33}$ values [5].

To provide a more useful understanding about the mechanical properties of the novel $C_6N_7$ monolayer, in Fig. 2 the uniaxial stress-strain responses are plotted for the loading along the armchair and zigzag directions (as distinguished in Fig. 1). It is clear that this monolayer shows isotropic elasticity and anisotropic tensile strength. Despite a nanoporous lattice, we predict remarkably high tensile strengths of 17.5 and 20.5 GPa, for the loading along the armchair and zigzag directions, respectively.

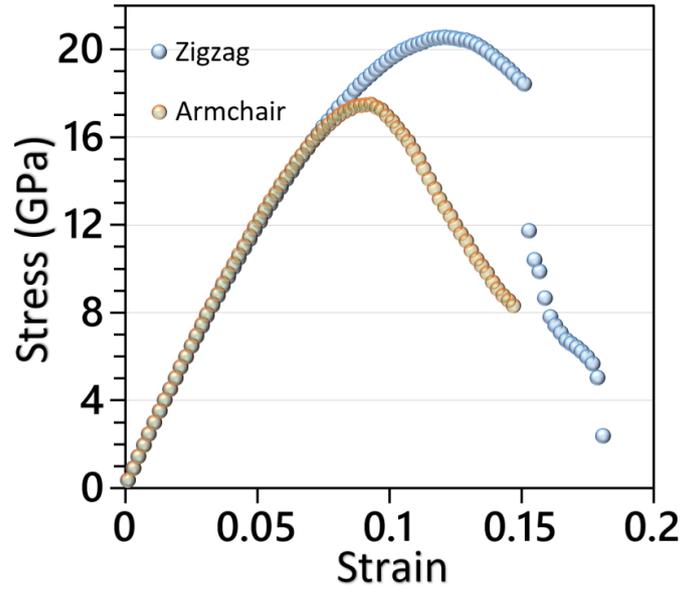

**Fig. 2**, Uniaxial stress-strain responses of the $C_6N_7$ monolayer along the armchair and zigzag directions.

In conclusion, the work by Bafekry *et al.*[4] includes significant errors in prediction of the elastic constants of the $C_6N_7$ monolayer. In contrast with their predicted very soft elastic modulus, we found that owing to formation of strong covalent C-C and C-N bonds throughout the carbon-nitride network, the $C_6N_7$ monolayer despite its porous structure is able to present a remarkably high elastic modulus of 267 GPa and a maximum tensile strength of 20.5 GPa.

**Acknowledgments**

B.M. appreciates the funding by the Deutsche Forschungsgemeinschaft (DFG, German Research Foundation) under Germany's Excellence Strategy within the Cluster of Excellence PhoenixD (EXC 2122, Project ID 390833453). F.S. thanks the Persian Gulf University Research Council, Iran for support of this study. B. M is greatly thankful to the VEGAS cluster at Bauhaus University of Weimar for providing the computational resources.